\documentclass[twocolumn]{aastex63}

\usepackage{comment} 
\received{June 1, 2019}
\revised{January 10, 2019}
\accepted{\today}

\submitjournal{AJ}

\shorttitle{Principal Component Analysis of X-ray Spectra}
\shortauthors{Rhea et al.}

\graphicspath{{./}{figures/}}
\usepackage{xcolor}
\definecolor{viridis1}{HTML}{FDE725}
\definecolor{viridis2}{HTML}{55C667}
\definecolor{viridis3}{HTML}{238A8D}
\definecolor{viridis4}{HTML}{440154}
\begin{document}

\title{A Novel Machine Learning Approach to Disentangle Multi-Temperature Regions in Galaxy Clusters}

\correspondingauthor{Carter Rhea}
\email{carterrhea@astro.umontreal}

\author[0000-0003-2001-1076]{Carter Rhea}
\affiliation{D\'epartement de Physique, Universit\'e de Montr\'eal, Succ. Centre-Ville, Montr\'eal, Qu\'ebec, H3C 3J7, Canada}

\author[0000-0001-7271-7340]{Julie Hlavacek-Larrondo}
\affiliation{D\'epartement de Physique, Universit\'e de Montr\'eal, Succ. Centre-Ville, Montr\'eal, Qu\'ebec, H3C 3J7, Canada}

\author[0000-0003-3544-3939]{Laurence Perreault-Levasseur}
\affiliation{D\'epartement de Physique, Universit\'e de Montr\'eal, Succ. Centre-Ville, Montr\'eal, Qu\'ebec, H3C 3J7, Canada}
\affiliation{Mila - Quebec Artificial Intelligence Institute, Montreal, Qu\'ebec, Canada}
\affiliation{Center for Computational Astrophysics, Flatiron Institute, New York, USA}

\author[0000-0002-7326-5793]{Marie-Lou Gendron-Marsolais}
\affiliation{European Southern Observatory, Alonso de Córdova 3107, Vitacura, Casilla 19001, Santiago, Chile}

\author[0000-0002-0765-0511]{Ralph Kraft}
\affiliation{Smithsonian Astrophysical Observatory, Cambridge, MA 02138, USA}

\begin{abstract}
The hot intra-cluster medium (ICM) surrounding the heart of galaxy clusters is a complex medium comprised of various emitting components. Although previous studies of nearby galaxy clusters, such as the Perseus, the Coma, or the Virgo cluster, have demonstrated the need for multiple thermal components when spectroscopically fitting the ICM's X-ray emission, no systematic methodology for calculating the number of underlying components currently exists. In turn, underestimating or overestimating the number of components can cause systematic errors in the emission parameter estimations.  
In this paper, we present a novel approach to determining the number of components using an amalgam of machine learning techniques. Synthetic spectra containing a various number of underlying thermal components were created using well-established tools available from the \textit{Chandra} X-ray Observatory. The dimensions of the training set was initially reduced using the Principal Component Analysis and then categorized based on the number of underlying components using a Random Forest Classifier. Our trained and tested algorithm was subsequently applied to \textit{Chandra} X-ray observations of the Perseus cluster. Our results demonstrate that machine learning techniques can efficiently and reliably estimate the number of underlying thermal components in the spectra of galaxy clusters, regardless of the thermal model (MEKAL versus APEC). 
We also confirm that the core of the Perseus cluster contains a mix of differing underlying thermal components. We emphasize that although this methodology was trained and applied on \textit{Chandra} X-ray observations, it is readily portable to other current (e.g. XMM-Newton, eROSITA) and upcoming (e.g. Athena, Lynx, XRISM) X-ray telescopes. The code is publicly available at \url{https://github.com/XtraAstronomy/Pumpkin}.
\end{abstract}


\section{Introduction} \label{sec:intro}
Galaxy clusters are massive structures that contain hundreds to thousands of galaxies. These environments accommodate extremely large reservoirs of hot gas ($\sim10^7-10^8$K) which constitute the intra-cluster medium (ICM; e.g. \citealt{sarazin_X-ray_1986}; \citealt{mushotzky_X-ray_1998}). Due to its elevated temperature, the ICM is a highly ionized plasma that emits primarily in the X-ray regime through the process of thermal bremmstrahlung (e.g. \citealt{markevitch_comparison_1997}; \citealt{ettori_coulomb_1998}; \citealt{sarazin_X-ray_1999}; \citealt{markevitch_temperature_1998}). While bremmstrahlung -- along with bound-free atomic transitions and the collisional excitation of hydrogen -- accounts for the continuum emission, several other mechanisms contribute to the total spectra. 
Many prominent emission lines, such as the Fe K-$\alpha$, Silicon, and Sulfer lines, are created through recombination and excitation mechanisms (e.g. see a review by \citealt{peterson_X-ray_2006}). Together, the continuum and descrete line emissions create a spectrum rich in information that can be studied to better understand the properties of hot astrophysical plasmas. 
 
The first X-ray observations of the ICM were taken fifty years ago of the Virgo, Perseus, and Coma clusters (\citealt{bradt_evidence_1967}; \citealt{gursky_strong_1971}; \citealt{gursky_X-ray_1973}). Over the following thirty years, the study of extragalactic X-ray spectra was developed through the use of orbital X-ray satellites such as \textit{Uhuru} (e.g. \citealt{sarazin_X-ray_1986}), the \textit{Einstein Observatory} (e.g. \citealt{forman_detection_1978}), \textit{EXOSAT} (e.g. \citealt{giacconi_einstein_1979}), and \textit{RXTE} (e.g.   \citealt{bradt_X-ray_1993}). Since the X-rays probe the hot plasma and are emitted primarily through thermal mechanisms, X-ray spectral analysis allows us to investigate the thermodynamic parameters of the ICM.
 Following the launch of the \textit{Chandra} X-ray Observatory and \textit{XMM-Newton}, our ability to map out thermodynamic properties improved dramatically due to the increased angular, spectral resolution and effective area (e.g. \citealt{wilman_fitting_1999}; \citealt{schindler_distant_1999}; \citealt{markevitch_chandra_2000}; \citealt{vikhlinin_evolution_2002}; \citealt{mazzotta_chandra_2001}; \citealt{forman_chandra_2002}; \citealt{forman_galaxy_2002}).
Recent surveys using these telescopes have revealed evolution in the temperature and pressure structure of galaxy clusters which hint at a change in the galactic environment over cosmological times while revealing a lack of evolution in the metalicity(e.g. \citealt{mcdonald_redshift_2014}; \citealt{cavagnolo_entropy_2008}; \citealt{bocquet_mass_2015}; \citealt{vikhlinin_evolution_2002}).
In addition to learning more about the cluster environment and its evolution, understanding the plasma's thermodynamic properties allow us to put stronger constraints on cosmology such as the galactic velocity dispersion - cluster X-ray mass, X-ray Luminosity - X-ray mass, and $\sigma_8$-$\Omega_m$ (e.g. \citealt{allen_cosmological_2003}; \citealt{allen_improved_2007}; \citealt{rosati_rosat_1997}; \citealt{horner_observational_1999}; \citealt{tozzi_evolution_2001}; \citealt{bocquet_mass_2015}). 

In order to access the parameters from the spectra (e.g. temperature, metallicity, pressure), researchers rely on fitting different thermal models which are dependent on the plasma's thermodynamics (e.g. \citealt{david_einstein_1990}; \citealt{markevitch_comparison_1997}; \citealt{markevitch_physics_1999}).
While several models have been developed to represent the X-ray emission of galaxy clusters, there are two predominant models in the literature: \texttt{APEC} (\citealt{smith_collisional_2001}) and \texttt{MEKAL} (\citealt{kaastra_X-ray_1993}). The models primarily differ in their background database of atomic line transitions which is used during the fitting procedure. Despite their differences, it is agreed that both work well and yield similar goodness-of-fit values (e.g. \citealt{brickhouse_notitle_2000}; \citealt{fabian_very_2006}; \citealt{sanders_deep_2010}). 
Additionally, recent observations have shown that different regions in a cluster may contain multiple components at different temperatures thus necessitating the use of several thermal models in the fitting procedure (e.g. \citealt{tamura_xmm-newton_2001}; \citealt{kaastra_spatially_2004}; \citealt{de_plaa_X-ray_2004}; \citealt{frank_characterization_2013} \citealt{boute_iron_2000}; \citealt{rasia_xmas2_2008}; \citealt{lovisari_non-uniformity_2019}).

 However, no systematic technique exists to determine the number of underlying thermal components in a given emission region. The current methodology requires users to fit multiple thermal components and accept that with the best reduced fit  (e.g. \citealt{churazov_xmmnewton_2003}; \citealt{fabian_very_2006}; \citealt{sanders_deeper_2007}; \citealt{fabian_wide_2011}; \citealt{zhuravleva_turbulent_2014}). While the generally accepted method -- fitting more components until the fit goodness-estimator is significantly reduced -- likely results in good estimates, the initial authors of the \texttt{APEC} model warn against using too many or too few components as doing so will skew the metallicity results; this is largely because the lower temperature components will dominate the metallicity measurement since they are more line driven than the hotter components (\citealt{raymond_soft_1977}; \citealt{smith_collisional_2001}). Additionally, obtaining an incorrect temperature estimate due to an incorrect number of assumed underlying components leads to systematic errors in all subsequently calculated values such as the cooling time and pressure. Moreover, determining the appropriate number of underlying thermal components necessary to model the multi-phase gas allows us to probe the underlying physics accurately (e.g. \citealt{sanders_deeper_2007}; \citealt{kaastra_spatially_2004}).

In this paper, we present a novel method for categorizing X-ray ICM emission through the use of two machine learning techniques: principal component analysis (PCA) and random forest classification. Using these techniques, we demonstrate they can be applied to emission spectra in order to classify the number of underlying thermal components. In $\S$ \ref{sec:meth}, we describe the PCA method and the creation of synthetic X-ray emission spectra. The primary results of this method are reported in $\S$ \ref{sec:results} in which we thoroughly test the algorithm on the synthetic spectra. A discussion of potential limitations of the algorithm are explored in $\S$ \ref{sec:disc}. In $\S$ \ref{sec:perseus}, the methodology is applied to \textit{Chandra} X-ray observations of the Perseus cluster. We discuss the impplications of this methodology to the larger X-ray galaxy cluster community in $\S$ \ref{sec:implications}. We also include a short discourse on the tutorials and software packaged created so that others can adapt our techniques for their own needs in $\S$ \ref{sec:conc}. Throughout this paper, we adopt a standard $\Lambda$CDM cosmology defined by $H_0=67\text{ km s}^{-1} \text{Mpc}^{-1}$ and $\Omega_M=0.3$. We have made the code public at \url{https://github.com/XtraAstronomy/Pumpkin}.

\section{Methodology and Data Reduction} \label{sec:meth}
\subsection{Principal Component Analysis}\label{sec:pca}
PCA is a popular statistical data reduction technique that breaks down complicated relations amongst variables into their primary components (e.g. \citealt{jolliffe_principal_2016}; \citealt{lever_principal_2017}; \citealt{bro_principal_2014}; \citealt{wold_principal_1987}; \citealt{shlens_tutorial_2014}). More precisely, PCA is a rotation of the data to a new, orthonormal basis in which the first coordinate contains a projection of the data which maximizes the variance (the first principal component), the second contains the second greatest variance (the second principal component), and so on. This technique is akin to calculating the eigenvalues/states which is ubiquitous in other multivariate statistical techniques such as Singular Value Decomposition (SVD) or Canonical Correlation Analysis (CCA). PCA has been used extensively in the scientific literature to create lower-dimensional representations of data by projecting the initial data into a subset of the PCA components. These components account for the majority of the variance of the data and retain the structures of interest. It is a linear transformation which is often used as a preprocessing to facilitate data classification or analysis. 
\begin{equation}\label{eq:pca}
    \vec{x_i} = \vec{\mu} + \sum_{j=1}^N a_{ij}\vec{v_j}
\end{equation}
Equation \ref{eq:pca} governs the principal component analysis (see appendix A for the mathematical details).
Since our work focuses on emission spectra, we describe the variables in terms of spectra.
$\vec{x_i}$ represents a given spectrum, $\vec{\mu}$ represents the dataset's mean spectrum, while each $a_{ij}\vec{v_j}$ represent the eigenvalue and eigenvector (eigen-spectrum) of the decomposition (\citealt{yip_spectral_2004}).  
Most relevant to our work, several authors have applied PCA to both stellar and galactic spectra (e.g. \citealt{ronen_principal_1999}; \citealt{mcgurk_principal_2010}; \citealt{pace_resolved_2019}). These works have demonstrated the success of machine learning techniques in extracting emission parameters from spectra. Here, our goal is to do the same for X-ray spectra of galaxy clusters.

\subsection{Decision Tree and Random Forest Classifier}\label{sec:DT-RFC}
In addition to PCA, we use the random forest classifier which builds upon the more fundamental decision tree model. These form a class of machine learning methods which are especially well-suited to classification tasks. In classification problems, data is fed into an algorithm to produce an output of interest (the class the input data belongs to). The parameters of this algorithm are learnt through a process called ‘training’, where the data for which the correct class is known are shown to the algorithm. During training, the values of these parameters are found by minimizing a distance, or ‘cost function’, between the output and the correct class.

We now discuss the training process in the specific case of the decision tree algorithm.
Starting with the root data set, the algorithm must first decide how to split the root node into sub-nodes. Generally, we employ a greedy algorithm that recursively calculates the cost of a split associated with each attribute in the data set (e.g. \citealt{quinlan_induction_1986}; \citealt{tan_introduction_2005}; \citealt{barros_survey_2012}). The algorithm then determines the split by taking the attribute-split that best minimizes the cost function. There are several cost functions available; however, they all have the same goal to create homogeneous branches (branches that have similar features). We use the Gini cost function which reduces the standard deviation within a proposed sub-node (\citealt{breiman_classification_1984}). Left unchecked, this process will create unwieldy trees that overfit the data. In order to inhibit this behavior, we either set a minimum number of inputs to be placed in each leaf and/or define a maximum recursion depth. Additionally, we can prune the tree and discard nodes that have minimal importance (e.g. \citealt{mingers_empirical_1989}; \citealt{song_decision_2015}). Although decision trees are easy to implement and useful in classification problems, they can suffer due to issues in the tree's variance or bias. One way to mitigate these effects is to build a random forest classifier.

Building upon the decision tree algorithm, a random forest classifier takes many individual decision trees and treats them as an ensemble. The guiding principal of the random forest classification is to create a large number of individual uncorrelated trees in order to make predictions. Classically, either a bagging -- bootstrap aggregation -- and/or a feature randomness algorithm are used to ensure the trees are relatively uncorrelated (e.g. \citealt{breiman_bagging_1996}; \citealt{breiman_random_2001}). For a detailed discussion of decision trees and random forest classifiers, we direct the reader to \cite{biau_analysis_2012}, \cite{denil_narrowing_2014}, and \cite{fawagreh_random_2014}.

In order to properly analyse the spectra of galaxy clusters and determine the underlying physical conditions of the hot gas, it is important to calculate how many temperature components are present in a given spectra. This problem naturally translates into a poly-modal classification problem that has been explored in the scientific community using machine learning algorithms. Typical classification algorithms include support vector machines, neural networks, and the decision tree -- and by extension random forest classifiers -- among others. Random forest classifiers have been used successfully in recent astronomy papers (e.g. \citealt{uzeirbegovic_eigengalaxies_2020}; \citealt{zhang_new_2019}; \citealt{beitia-antero_use_2018}). 
Therefore, we opted to utilize the standard random forest algorithm implemented in \texttt{SKLEARN}.

Although the random forest algorithm is designed to handle a single input, we must adapt it since we will often have multiple spectra of the same region. This occurs when exposures are taken at different epochs -- a technique commonly employed in observational astronomy. In order to apply a random forest classifier to several spectra at once, we apply the classifier to each spectrum individually. We then leverage the definition of the classifier and sum the probabilities over the ensemble. The final classification is the class with the highest summed probability. Since each spectrum's classifier is independent, there is no notion of "simultaneous" adjustment. The author's note that having a simultaneous random forest classifier (also known as a Joint Random Forest Classifier) could be preferable; the technique is currently under development (e.g. \citealt{petralia_new_2016}).

\subsection{Synthetic Chandra Spectra}\label{sec:synchan}
Since our ultimate goal is to devise a systematic method to determine the number of underlying temperature components in a given X-ray spectrum, we must train our algorithm with emission spectra that contain multiple temperature components. Since this value is unknown for real observations, we construct a well-rounded set of synthetic spectra with differing emission parameters. Moreover, we choose to create synthetic \textit{Chandra} X-ray Observatory spectra because of the telescopes un-paralleled spatial resolution which will allow us to probe smaller regions in which the number of temperature components may change. Additionally, \textit{Chandra}'s spectral resolution provides an adequate number of emission features to train the algorithm. Although \textit{Chandra} was chosen for this study, our application can be ported to other missions such as \textit{XMM-Newton} and eventually \textit{Athena}. Finally, several observations of nearby galaxy clusters with a complex temperature structure, such as M87, Perseus, and Coma, have been completed by \textit{Chandra}.

Synthetic spectra were created using \texttt{SHERPA's FAKE\_PHA} tool. 
The tool requires the use of a chosen response matrix file (rmf) and ancillary response file (arf). The \textit{Chandra} detectors have been steadily degrading each year which has lead to a consitent change in the reponse matrix. Since the observations used later in this article (see $\S$\ref{sec:perseus}) were taken during cycle 03, we use an rmf and arf file from the same epoch. Following a discussion with the \textit{Chandra} X-ray Obersvatory Helpdesk (private communications), we adopt the responses matrices from one of the observations of the Perseus cluster explored in later sections: ObsID 3209. The rmf and arf files were created using the \texttt{specextract} tool. The region over which we calculated the response matrix is defined in $\S$\ref{sec:perseus}.  We note that, for the present moment, in order to extend this work to other epochs, the synthetic spectra must be reconstructed and thus the algorithms must be retrained. In a future paper, we will explore the variations of responses matrices further.

We constructed temperature emission profiles with various numbers of underlying thermal components using an absorbed thermal emission model taken from the \texttt{XSPEC} package: \texttt{PHABS*APEC},  \texttt{PHABS*(APEC1+APEC2)},
\texttt{PHABS*(APEC1+APEC2+APEC3)}, \texttt{PHABS*(APEC1+APEC2+APEC3+APEC4)} where \texttt{PHABS} absorption represents galactic absorption. We use \texttt{APEC v3.0.9} in order to model the thermal emission. We chose not to include more than 4 thermal components since most of the literature does not include more (e.g. \citealt{fabian_very_2006}; \citealt{sanders_deeper_2007}). We created 25,000 spectra for each number of underlying thermal components; thus, we have a total of 100,000 spectra. In order to avoid any potential bias in the ordering of the spectra, we randomly selected with replacement 100,000 spectra from our sample, meaning we potentially select the same spectrum more than once; this commonly used method is known as bootstrapping. Allowing 90\% of our synthetic spectra to be placed in the training set, we used the remaining 10\% for the test set. This division of data is standard in machine learning applications (e.g. \citealt{breiman_random_2001}). We run a random grid cross-validation (RGCV) search on the training set to tune the random forest hyper-parameters. At test time, the hyper-parameters are static and set to the optimal results from the RGCV search.

The algorithm outline is as follows:
\begin{enumerate}
    \item Construct synthetic spectra for single-, double-, triple-, and quadruple-temperature models.
    \item Create co-variance/projection matrices from PCA analysis on the training set.
    \item Train the random forest algorithm using the principal components of our synthetic data to classify spectra by their number of underlying thermal components.
    \item Project the test set into the principal component basis and verify the viability of the trained algorithm on the projected test set.
\end{enumerate}

In order to verify that our algorithm works with real observations, we use \textit{Chandra} observations of the Perseus cluster. Perseus was chosen since it is a massive, nearby cool-core galaxy cluster for which there exists a vast literature on the X-ray emission (e.g. \citealt{fabian_deep_2003}; \citealt{fabian_very_2006}; \citealt{fabian_wide_2011}; \citealt{sanders_deeper_2007}). Although we chose the Perseus cluster, our methodology could be readily tested on other nearby galaxy clusters such as  Coma or Virgo. With a test cluster defined, we select emission parameters which coincide with the Perseus cluster.
Unless otherwise noted, we adopted a column density, $n_H$ = 0.14$\times10^{22}$cm$^2$ (\citealt{kalberla_leidenargentinebonn_2005}), and a redshift equal to 0.018, chosen to coincide with the Perseus cluster (e.g. \citealt{gudehus_systematic_1991}; \citealt{hudson_galaxy_1997}; \citealt{hicken_improved_2009}). In order to demonstrate the feasibility of the algorithm for any nearby galaxy cluster, we also tested lower redshift values: 0.01 and 0.005. Equivalently, the chosen column density closely corresponds to that in the direction of the Perseus cluster (\citealt{kalberla_leidenargentinebonn_2005}). The temperature values were randomly sampled between $0.1-4.0$ keV since the majority of nearby galaxy cluster thermal emission is within this range, the range contains both soft and hard emission lines which are critical in fitting procedures, and is well within the observing band of \textit{Chandra} (e.g. \citealt{henriksen_physical_1985}; \citealt{fabian_deep_2003}; \citealt{peterson_X-ray_2006}; \citealt{bohringer_X-ray_2010}; \citealt{mushotzky_X-ray_1984}; \citealt{mohr_properties_1999}; \citealt{loewenstein_chemical_2003}; \citealt{odell_radiation_2000}). No minimum temperature separation between componants was imposed. The ramifications of this are explored in section $\S$ \ref{sec:mincomp}. We also varied the metallicity between $0.2-1.0$ Z$_\odot$  -- a standard range adopted for ICM (e.g. \citealt{allen_relationship_1998}; \citealt{mushotzky_X-ray_1998}; \citealt{peterson_X-ray_2006}). We choose a target signal-to-noise value equal to 150 or approximately 22,000 counts. \deleted{Although this value is high, we want the algorithm to learn to recognize the most important emission lines. Moreover, }This value coincides with that used in studies of the Perseus cluster (e.g. \citealt{fabian_deep_2003}; \citealt{fabian_very_2006}; \citealt{sanders_deeper_2007}). In order to demonstrate the robustness of the algorithm to lower signal-to-noise values, we also tested its performance on data with a signal-to-noise value of 50 (2500 counts) in $\S$ \ref{sec:StN}. 

\subsection{Chandra observations of the Perseus cluster }\label{sec:Perseus_clean}
The Perseus cluster has been observed for over 1.4 mega seconds with the \textit{Chandra} X-ray Observatory. Although the majority of the observations focus on the cluster's core (which is believed to contain multi-temperature regions according to, for example, \citealt{sanders_deep_2010}), we select ObsID 3209 (81.4 ks) and ObsID 4289 (90.4 ks). These two observations were selected for several reasons: they were taken at approximately the same time and each contain a significant number of counts. In both observations the front-illuminated ACIS-S2, ACIS-I1, and ACIS-I3 chips were activated. 
Starting with the level I event file provided by the \textit{Chandra} X-ray Observatory (CXC) team, we  followed a standard cleaning and reduction technique using \texttt{CIAO V4.12, CALDB V4.9.0, PYTHON V3.5.1}. We first removed point-sources detected by \texttt{VTPDETECT} from a background CCD (ACIS-I1) and then applied the \texttt{LC\_SIGMA\_CLIP} script with a 2$\sigma$ threshold to removed time intervals exhibiting background flares. We proceeded to apply time-dependent and charge-transfer gain corrections, destreak, and process the data using the \texttt{CHANDRA\_REPRO} tool with \texttt{VFAINT=TRUE}. 
An exposure corrected, background subtracted, merged image between 0.5-7.0 keV was created using the \texttt{merge\_obs} tool. Spectroscopic fits using \texttt{SHERPA V4.12.0} and \texttt{XSPEC V12.10.1} are described in $\S$\ref{sec:perseus}.

\section{Results on the synthetic spectra} \label{sec:results}
Before applying our methodology to real data, we tested it against synthetic data. These results lay the foundation of our future results.

\subsection{Principal Component Analysis}\label{sec:resultsPCA}

\begin{figure}[h!]
    \centering
    \includegraphics[width=0.5\textwidth]{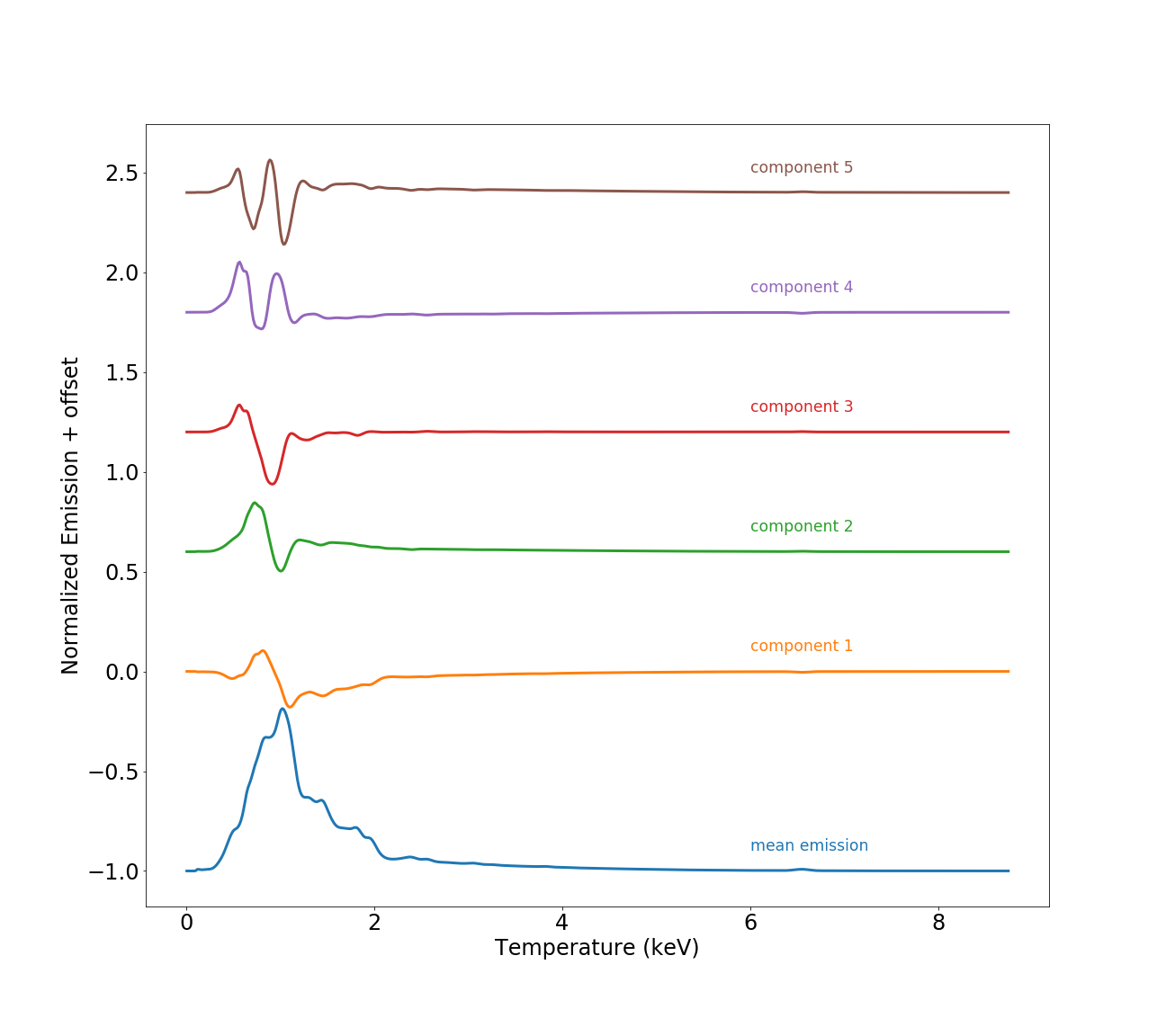}
    \caption{The Principal Component Analysis (PCA) effectively reduces a higher dimensional problem into a linear combination of eigen-vectors (eigen-spectra) and a mean emission profile -- see equation \ref{eq:pca}. In this graphic, we visualize the mean emission spectra, as determined by the PCA, and the first five eigen-spectra.}
    \label{fig:PCA}
\end{figure}
We apply the \texttt{SKLEARN} implementation of the PCA method as described in $\S$\ref{sec:pca} to the 36000 synthetic spectra that comprise our training set. Figure \ref{fig:PCA} shows the mean spectra and the eigen-spectra of our training set. The eigen-spectra are the principal components projected onto the original domain (they are the $\vec{v_j}$ in equation \ref{eq:pca}).  We note that the primary variations in all components are visible in the soft X-ray regime (0.5-2.0 keV). This trend is expected since we are primarily modeling the diffuse hot gas which emits mainly in this regime. The first two eigen-spectra can be interpreted physically as follows.
The first eigen-spectra, component 1 in Figure \ref{fig:PCA}, captures the difference in the ultra soft (0.5-1.0 keV) and soft diffuse emission (1.0-2.0 keV). The second eigen-spectra, component 2 in Figure \ref{fig:PCA}, captures the Fe-L/Ne and Mg emission lines. The remaining principal components do not have a clear physical interpretation. This is often an issue in principal component analysis (e.g. \citealt{jolliffe_principal_2016}).

In order to determine how many principal components we need to include, we study the overall variance retained versus the number of principal components. We find that, in order to capture $99\%$ of the variance in our synthetic data, 25 principal components are required. After 25 principal components, the variance plateaus. Hence, we adopt 25 as the optimal number of principal components.

\begin{figure*}
\gridline{\fig{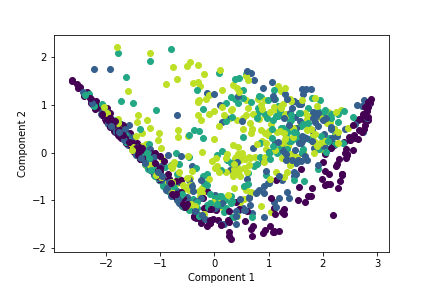}{0.3\textwidth}{(a)}
          \fig{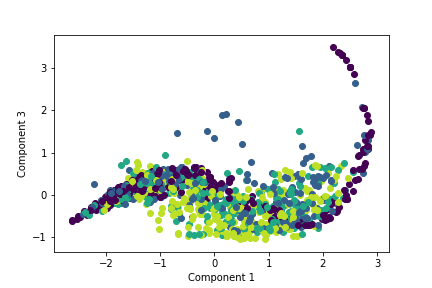}{0.3\textwidth}{(b)}
          \fig{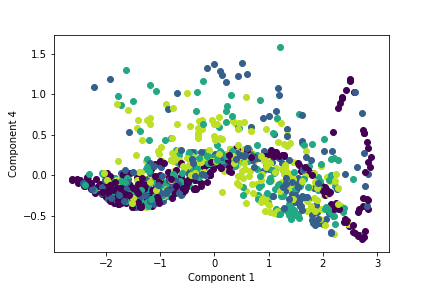}{0.3\textwidth}{(c)}
          }
\gridline{\fig{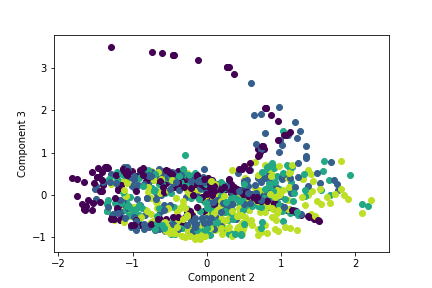}{0.3\textwidth}{(d)}
          \fig{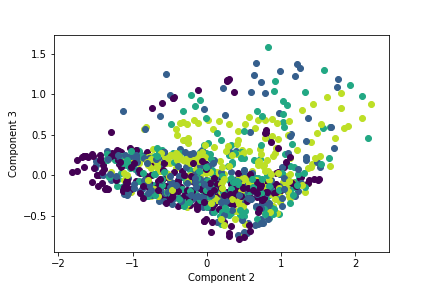}{0.3\textwidth}{(e)}
          \fig{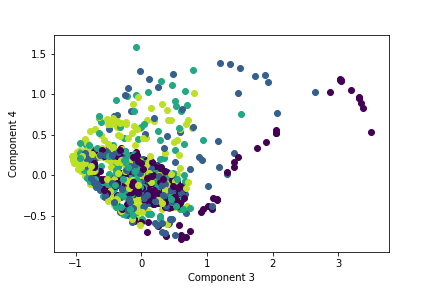}{0.3\textwidth}{(f)}
          }

\caption{Principal Component Analysis (PCA) Comparison Plots for 1000 randomly selected training spectra. The number of underlying thermal components are color coded in the following fashion: {\color{viridis1}Single}, {\color{viridis2}Double}, {\color{viridis3}Triple}, {\color{viridis4}Quadruple}. (a) Component 1 vs 2, (b) Component 1 vs 3, (c) Component 1 vs 4, (d) Component 2 vs 3, (e) Component 2 vs 4, (f) Component 3 vs 4.}
\label{fig:PCA_components}
\end{figure*}

Figure \ref{fig:PCA_components} visualizes the projection of our training set (36000 spectra) onto the principal component basis (the $a_{ij}$ values in equation \ref{eq:pca}). The principal component vectors represent the eigen-spectra which can be added linearly to reconstruct 99\% of the variance in a given spectrum. We will use these components as the inputs for our random forest classifier since they capture the variance in the spectra in a lower-dimensional set compared to the un-adulterated spectra. The random forest classifier will be used to determine the number of underlying temperature components in a spectra given its principal components.

Although the data does not segregate itself into different regions in each component, we can see that certain trends exists for the different temperature bins which allow for easier classification in later stages of our methodology. For example, in Figure \ref{fig:PCA_components} (a), spectra with four components follow a tight negative trend until component one is approximately 3 and then abruptly changes to a positive trend. Thus, if the first and second principal components of a spectrum do not lie on this trend, the spectra will not be classified as having four underlying thermal components.

\subsection{Random Forest Classification}\label{sec:ml_basic}
 Once the projection matrix is calculated from the spectra in the training set, we apply it to obtain the 25 first principal components of every spectra in the training set, and use those to train our random forest classifier. 
\begin{figure}[!h]
    \centering
    \includegraphics[width=0.5\textwidth]{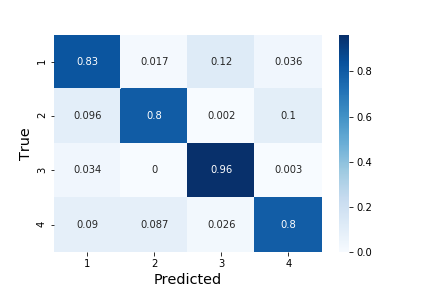}
    \caption{Random forest confusion matrix evaluated on the test set consisting of 8,000 spectra. The X and Y axis represent the number of underlying thermal components predicted and actually present in the spectra.  The majority of counts are bundled along the diagonal, indicating the relatively-high predictive accuracy of the method.  Values are normalized by the total number of counts per row.}
    \label{fig:PCA_ML}
\end{figure}
After the training of our decision tree, we will be able to apply it to other spectra by using our pre-calculated projection matrix to reproject the new spectrum into the principal component basis and applying the pre-trained random forest classification algorithm. We define training to be successful if we can achieve higher than 50\% accuracy in our test set. We use confusion matrices to validate the results; confusion matrices are frequently used in the machine learning community to visualize the accuracy of the predictions against a validation set (\citealt{kohavi_applied_1998}). An optimal algorithm will have all the values along the diagonal -- this indicates that the algorithm's predictions match the ground truth values which are in our case the number of underlying spectral components.

\begin{figure}[!h]
    \centering
    \includegraphics[width=0.5\textwidth]{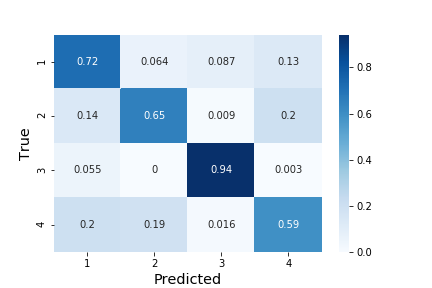}
    \caption{Confusion Matrix created by omitting the principal component analysis and running the random forest directly on the spectra themselves.}
    \label{fig:ML_noPCA}
\end{figure}

Figure \ref{fig:PCA_ML} demonstrates the efficacy of this model for estimating the number of thermal components in synthetic X-ray spectra since the matrix is mainly diagonal. A diagonal covariance matrix indicates that the method is properly categorizing the number of underlying thermal components. For example, the algorithm correctly predicts that a spectrum has three underlying components 95\% of the time. The third row also indicates that the algorithm incorrectly predicts the three-component spectrum as having one component 4.6\%, two components 0.1\%, and four components 0.7\% of the time.
The graphic also notes the weak points in the algorithm: primarily distinguishing between two and four thermal components. 
We repeated the same machine learning algorithm neglecting the PCA step (i.e. running our random forest directly on the spectra). A comparison of Figure \ref{fig:PCA_ML} (PCA) and Figure \ref{fig:ML_noPCA} (no PCA) reveals the power of using principal component analysis to extract the crucial spectral features.

\section{Discussion}\label{sec:disc}
\subsection{Limitations}
As with the development of any new method, it is important to understand its limitations. The following section explores many ways in which we attempted to "break" our algorithm and the resulting conclusions.

\subsubsection{Application to More Components}
\begin{figure}[!ht]
    \centering
    \includegraphics[width=0.45\textwidth]{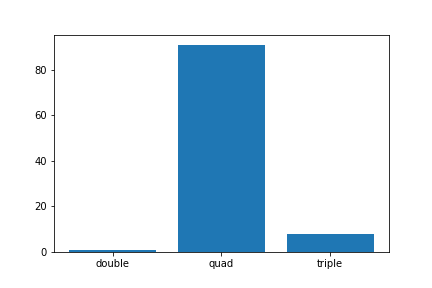}
    \caption{Using our 100 spectra with five underlying thermal components, we applied our methodology. On the y-axis, we can see the number of spectra categorized by the algorithm as either double (2 components), triple (3 components), or quad (4 components). }
    \label{fig:Pent}
\end{figure}

In order to determine how the methodology handles regions with more than four underlying thermal components, we created 100 synthetic spectra with five thermal components: \texttt{PHABS*(APEC1+APEC2+APEC3+APEC4+APEC5)}. We then applied our methodology to this set. As readily seen in figure \ref{fig:Pent}, the algorithm heavily favors a four-component model.

This indicates that the algorithm is able to accurately realize that there are at least four thermal components. As such, we recommend using a classification as four-component as a lower limit for the number of underlying thermal components. While beyond the scope of this paper, it would be worthwhile to explore an algorithm extending up to higher numbers of components. We strongly suggest users do not use the random forest trained as such to classify more than four components since the extrapolation abilities of such methods are limited.

\subsubsection{Model Dependence: \texttt{APEC} vs. \texttt{MEKAL}}
\begin{figure}[!h]
    \centering
    \includegraphics[width=0.45\textwidth]{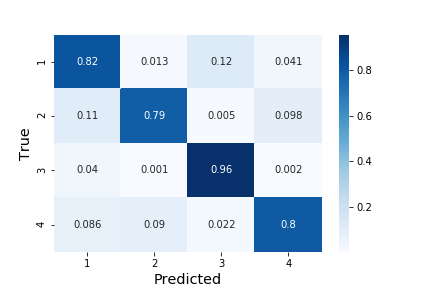}
    \caption{Confusion matrix for an algorithm trained and verified on synthetic spectra generated using the \texttt{MEKAL} thermal model. We note consistent results regardless of the training set's thermal model.}
    \label{fig:PCA_Mekal}
\end{figure}

As noted, we created our synthetic spectra using the \texttt{APEC} model. However, the observational X-ray astronomy community still debates the advantages of the \texttt{APEC} model over its predecessor, the \texttt{MEKAL} model. In order to demonstrate that our methodology is thermal-model agnostic, we asked two questions: does the algorithm work as well for synthetic spectra built from the \texttt{MEKAL} model and can the algorithm accurately classify \texttt{MEKAL} spectra if trained on an \texttt{APEC} model and vice-versa? We once again created 10,000 synthetic spectra for single, double, triple, and quadruple temperature components (totaling 40,000 spectra), but this time using the \texttt{MEKAL} model: \texttt{PHABS*MEKAL}. We use the \texttt{MEKAL} model realization in the \texttt{CIAO} package. We then trained and tested our algorithm in an identical manner. Figure \ref{fig:PCA_Mekal} demonstrates that the random forest performs equally as well for the \texttt{MEKAL} model as the \texttt{APEC} model. Additionally, we tested if an algorithm trained on data created using the \texttt{APEC} model could accurately categorize data created using the \texttt{MEKAL} model and vice-versa. Unfortunately, under these conditions, our algorithm does slightly better than random choice. Thus, it is important that, for real observations, the \texttt{APEC}-trained and \texttt{MEKAL}-trained models return the same predictions since it is still debated whether the \texttt{APEC} or \texttt{MEKAL} model is more appropriate. 

\subsubsection{Signal to Noise Constraints}\label{sec:StN}
\begin{figure}[!hb]
    \centering
    \includegraphics[width=0.45\textwidth]{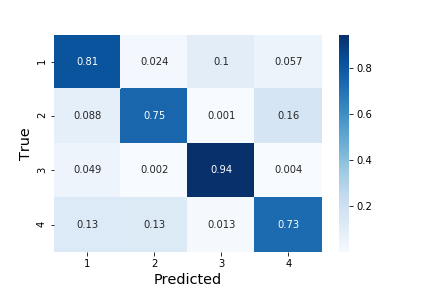} 
    \caption{We used the algorithm trained on a signal-to-noise ratio of 150 and applied it to spectra with a signal-to-noise ratio of 50. Though the random forest loses some of its predictive power, it still estimates the correct number of parameters the majority of the time in all cases. This is likely due to the effects of the increased noise. }%
    \label{fig:lowCts}%
\end{figure}

Our initial set of synthetic spectra were created assuming a signal-to-noise ratio of 150. Here, we explore the dependence of our algorithm predictive powers on the signal-to-noise of the test spectra. Using our pre-trained algorithm, we predicted the number of thermal-emission components in synthetic spectra with a signal-to-noise ratio of 50. A signal-to-noise value of 50 was chosen since it is a commonly chosen value in the literature because it ensures that the spectra will have enough data to constrain the thermodynamic parameters (e.g. \citealt{diehl_adaptive_2006}; \citealt{schenck_X-ray_2014}; \citealt{datta_how_2014}). We can see from figure \ref{fig:lowCts} that the algorithm continues to perform well on lower signal-to-noise data than its initial training set. This is important because it allows us to use this technique on exposures that are shallower than that of the Perseus cluster.

\subsubsection{Relative Abundance of Thermal Components}
An often overlooked question when fitting multiple thermal components is the relative strength of the different components. Physically this could be due to either the relative density differences or amount of the plasma in that state in a given region. Relative strengths of the components were randomly varied and normalized to the sum to 1\footnote{As an example, if we have a spectra with two underlying components, we allow the strength of the first thermal component to be 30\% and the strength of the second component to be 70\%.}. We created 15,000 synthetic spectra from each number of temperature components resulting in a total of 60,000 synthetic spectra with varying relative strengths. 
No changes were seen in the confusion matrix, and so we note that changes to the relative strengths of thermal components does not  affect the reliability of our methodology. We note, however, that it is highly unlikely for the network to correctly classify a 2T plasma for which one component contributes 1\% and the other 99\%. Throughout the remainder of the paper, we make the assumption that each thermal component contributes equally to the overall spectrum.

\subsection{Minimal Separation between Thermal Components}\label{sec:mincomp}

As mentioned in $\S$ \ref{sec:synchan}, we imposed no minimum separation of the temperature of each component in spectra with multiple underlying thermal components. In order to asses whether or not this negatively affected our classifications, we studied the spectra that were mis-classified by the random forest. If the lack of a minimum separation was responsible for the mis-classifications, we would expect the mean separation in temperatures of the different components in the mis-classified spectra to be small. When calculated, we find the mean temperature separation between components to be approximately 1.3 keV (with a variance of approximately 0.3 keV) which is consistent given the uniform distribution over the chosen range of temperatures (0.1 - 4.0 keV) used to construct the spectra. We conclude that mis-classifications are not due to components having near-identical temperatures in the training set; therefore, the accuracy of the method would not benefit from imposing a minimum separation between thermal components. The well-classified data has approximately the same mean separation and mean variance in the separation. Morever, the goal of this work is not to distinguish between minimally separated thermal components, but rather to determine the number of primary thermal components necessary to describe the emission.

\subsection{Necessity and Meaning of Multiple Components}

As outlined in $\S$\ref{sec:intro}, it is critical that the appropriate number of underlying thermal components is chosen so that the physics of the system can be understood. Although the X-ray spectra is undoubtedly a results of a continuum of thermal emission components, there exist peaks in this distribution (e.g. \citealt{kaastra_spatially_2004}). When we apply procedures to search for multi-phase gas, we are indeed searching for the temperature peaks in the distribution. In addition for the need to categorize the gas correctly in order to accurately describe the plasma's physics, underestimating the number of underlying thermal components has implications for the efficacy of fits. In order to quantify these implications, we consider a plasma with two underlying components. We generate a mock \textit{Chandra} spectrum assuming a SNR of 150 which is comprised of two \texttt{APEC} models at redshift z=0 with $Z_{met}=0.3$M$_\odot$ with temperatures set to 4 keV and 8 keV. We bin the data at 50 counts per bin. We attempt to fit a single \texttt{APEC} model to the mock spectra. The model finds a best-fit temperature of 5.3 keV. However, the fit statistic ($\chi^2 = 2.3$) and q-value ($q\sim10^{-42}$) indicate that the fit is unacceptable. Thus we are confident that a single thermal component does not adequately describe a spectrum with two underlying thermal components.

\section{Perseus cluster Observations}\label{sec:perseus}

\begin{figure*}[]
    \centering
    \plotone{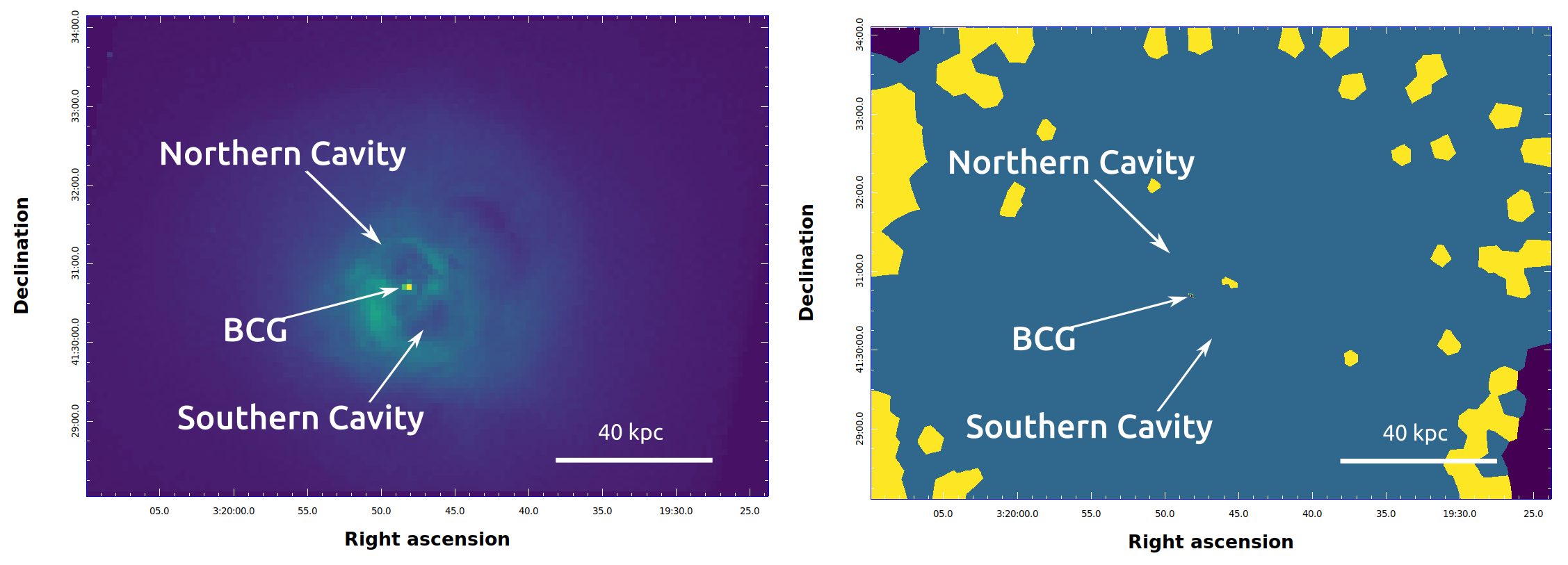}
    \caption{
    Left: Background-subtracted, exposure-corrected, and smoothed image of the core of the Perseus cluster. The emission shown is between 0.1-8 keV.
    Right: A weighted voronoi tessellation map of the Perseus cluster using a signal-to-noise value of 150. We have a total of 1271 bins. The bins are color-coded in order to reveal the number of underlying temperature-components hidden in the spectra as classified by our random forest algorithm in the following manner: {\color{viridis4}Single}, {\color{viridis3}Double}, {\color{viridis2}Triple}, {\color{viridis1}Quadruple}. The image clearly demonstrates the necessity of multiple components when modeling the hot plasma's X-ray emission captured by \textit{Chandra}.}
    \label{fig:Persues_ab}
\end{figure*}

In Fig. \ref{fig:Persues_ab}, we show the Chandra X-ray observations of the Perseus cluster used to test our machine learning approach. Our goal is to apply our techniques outlined in Section \ref{sec:meth} and determine the number of thermal emission components in a given region of the Perseus cluster. To do this, we first applied the weighted voronoi tessellation (WVT) algorithm to the reduced X-ray data described in section $\S$\ref{sec:Perseus_clean}. The bin map resulting from the WVT algorithm is routinely used to study the structure of ICM X-ray emission (e.g. \citealt{cappellari_adaptive_2003}; \citealt{diehl_adaptive_2006}). The WVT algorithm works by binning the merged image into self-similar signal-to-noise regions initially using a simple bin accretion algorithm. These regions are used as an initial guess for the true WVT algorithm which minimizes a scale length, defined \textit{a-priori}, so that the pixels are grouped into final, optimized bins. Thus, the final product is an image of binned pixels which indicate the underlying signal-to-noise structure\footnote{Our implementation of this technique can be found at \href{https://github.com/crhea93/AstronomyTools}{https://github.com/crhea93/AstronomyTools} under the Weighted Voronoi Tessellation directory which contains further documentation and testing.}. Setting the target signal-to-noise ratio of 150 (22,000 counts) results in 916 binned regions for the cluster core (see magenta box in Fig. \ref{fig:Persues_ab}). 
Having completed the binning map, we proceeded to create a corresponding spectrum for each region using the \texttt{CIAO} tool \texttt{SPECEXTRACT} for each ObsID; additionally, we use a background region created from a blanksky file created with the \texttt{BLANKSKY} script. 

\begin{figure*}[]
    \centering
    \plotone{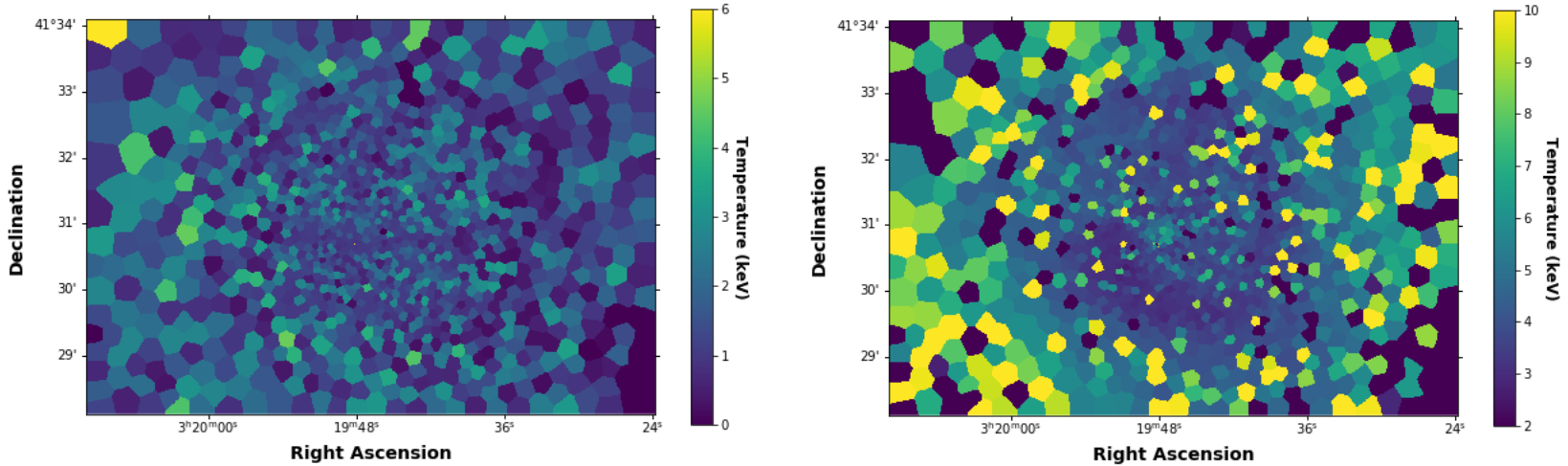}
    \caption{Temperature Maps of the Perseus cluster in regions designated as containing two or more underlying thermal components. The first component is shown on the left, while the second component is shown on the right. The figure demonstrates that the two temperature components are well separated. We only show regions for which the fits recovered acceptable reduced chi statistics ($0.8<\chi^2<1.2$).
    }
    \label{fig:Persues_temps}
\end{figure*}

Using the spectral files for each region, we applied our methodology as follows. We first employed our principal component analysis decomposition on each combined spectrum in order to project the spectrum to the PCA-space. We then used our trained\footnote{We trained the algorithm twice: once with the \texttt{APEC} model and once with the \texttt{MEKAL} model. We report that both algorithms result in the same underlying temperature map.} random forest classifier to predict the number of underlying thermal components in each region's spectrum. This was done for both ObsIDs: 3209 and 4289. We then selected the final classification as the output with the highest aggregate probablity after combining the results of the two independent classifiers (see $\S$ \ref{sec:DT-RFC}). 
The results can be seen in Figure \ref{fig:Persues_ab}. As suspected, the ICM in the Perseus cluster cannot be classified as a single-temperature plasma, but rather, our methodology has revealed that several thermal components are necessary to properly model it; this is congruent with studies using traditional methods (\citealt{fabian_very_2006}; \citealt{sanders_deeper_2007}; \citealt{fabian_wide_2011}). Moreover, the algorithm reveals that the majority of the ICM has two underlying components. A map showing the confidence with which each region is assigned a number of underlying thermal components is shown in appendix C.

Additionally we calculate temperature maps for the regions containing two underlying thermal components. Temperature maps were calculated following standard fitting procedures outlined by the \textit{Chandra} X-ray Observatory (e.g. \citealt{fabian_very_2006}; \citealt{sanders_mapping_2004}). We use \texttt{Sherpa v4.12.1} to simultaneously fit the spectra of regions defined by our weighted voronoi tessellation map for each observation with an absorbed thermal model (\texttt{PHABS*APEC}). Similar to the procedure in $\S$ \ref{sec:synchan}, we adopted a redshift of 0.0179 and column density,$n_H$, equal to 0.14$\times10^{22}$cm$^2$. We model the background X-ray emission with a soft X-ray Galactic component ({\sc{apec}}) with a temperature of $0.18$keV and metallicity Z=1 and a hard cosmic X-ray component ({\sc{bremss}}) with a temperature of $kT=40$keV. The data were binned at 50 counts. The temperature maps for each component are shown in figure \ref{fig:Persues_temps}. Similar to the results show in figure 12 of \cite{sanders_mapping_2004}, we find that the two principal thermal components in the central regions of the cluster are characterized by a cooler ($\approx 2$keV) and hotter ($\approx 4$keV) gas.

While we do not explore in detail the thermodynamic properties of the Perseus cluster in this paper, we have  demonstrated not only the applicability, but also the virtue and potency, of our methodology when adapted to real observations. 
We also note that we do not compare the number of underlying thermal components determined by our algorithm to the number expected using the standard method since the standard method contains inherent issues.

\section{Implications for the X-ray community}\label{sec:implications} 

We emphasize that despite the fact that this article focuses on data obtained by the \textit{Chandra} X-ray Observatory, this technique can be applied to other X-ray missions such as \textit{XMM-Newton}, \textit{Athena} or \textit{XRISM}. Moreover, this technique will allow for the fast and unbiased -- in the sense that it requires no human intervention -- categorization of galaxy cluster spectra which will be important in upcoming X-ray survey missions such as eROSITA (e.g. \citealt{merloni_erosita_2012}). With the advent of X-ray telescopes that have a higher spectral resolution, such as the Athena Space X-ray Observatory (e.g. \citealt{barret_athena_2020}), we can expect this classification methodology to perform even more accurately since there will be more emission features to use in the classification. However, a change in either the spectral resolution or instrument will require retraining the algorithm. Doing so will ensure that the algorithm is learning the proper response matrices and guiding spectral lines/ratios. We also note that, with the proper training set, this methodology can be applied to other astrophysical phenomenon that emit in the X-rays such as AGN and supernovae remnants. 

Although the algorithm performs well on test data and real observations, an additional complication arises when applying the algorithm to a different time epoch (i.e. a \textit{Chandra} cycle other than cycle 03). As discussed, the CCD cameras have been degrading over time, thus the response matrices change from cycle to cycle. In a future paper, we will explore different techniques to design the machine learning methodology presented here cycle-agnostic. In a separate follow-up paper, we will also explore the applicability of machine learning methods to the prediction of spectral emission parameters such as temperature and metallicity using \textit{Chandra} X-ray spectra.

\section{Conclusions} \label{sec:conc}
This paper has explored the efficacy of principal component analysis coupled with a random forest classifier for classifying the number of underlying thermal-emission components in the X-ray spectra of hot gas in galaxy clusters. 
The python code package, Pumpkin, created for this analysis is readily available at the following github address: \url{https://github.com/XtraAstronomy/Pumpkin}.

We have included several examples in the form of jupyter notebooks in order to facilitate the reproducability of our results and to make our code more accessible to the community. To address potential issues regarding our training set (i.e. choice of redshift, column density, and temperature range), we have included a tutorial on creating synthetic spectra and training the random forest algorithm so that our methodology can be easily applied to galaxy clusters and groups at different epochs. We have also included a tutorial on applying the pre-trained PCA and random forest algorithms directly to observations.

Our primary conclusions are as follows:
\begin{itemize}
	\item 
	We report the success of our methodology in estimating the number of thermal components on both synthetic and real X-ray observations of galaxy clusters. A comparison with the literature revealed concurrent results in the case of the Perseus cluster.
	\item 
	We explored the effects of different temperature models, \texttt{MEKAL} and \texttt{APEC}, on the algorithm. There are no discernible effects dependant on the chosen model.  However, the algorithm does not reliably predict the number of underlying thermal components if trained on one temperature model then tested on the other.
	\item 
	We note that a slight decrease in signal-to-noise (150 to 50) does not drastically affect the accuracy of the predictiosns 
	\item 
	The redshift and column density were uniformly sampled from a realistic range in order to demonstrate that the algorithm is not negatively affected by natural heterogeneity in the parameter space. 
	\item 
	We confirm that the core of the Perseus cluster is best categorized by several thermal emission components rather than by a single component.
	\item 
	We developed several tutorials for the use and adaptability of our algorithm.
\end{itemize}
The following paper in this series will focus on the feasibility of using machine learning to better understand the temperature parameter of X-ray spectra.

\acknowledgments
C. R. acknowledges financial support from the physics department of the Universit\'e de Montr\'eal.
J. H.-L. acknowledges support from NSERC via the Discovery grant program, as well as the Canada Research Chair program.
We would like to thank Farbod Jarandar for the enlightening discussions on machine learning techniques for spectral analysis.

The progamming aspects of this paper were completed thanks to the following packages of the python programming language (\citealt{van_rossum_python_2009}): \texttt{numpy} (\citealt{van_der_walt_numpy_2011}, \texttt{scipy} (\citealt{virtanen_scipy_2020}), \texttt{matplotlib} (\citealt{hunter_matplotlib_2007}), \texttt{pandas} (\citealt{mckinney_data_2010}), \texttt{seaborn} (\citealt{michael_waskom_mwaskomseaborn_2017}), \texttt{astropy} (\citealt{the_astropy_collaboration_astropy_2018}; \citealt{robitaille_astropy_2013}), \texttt{sherpa} (\citealt{freeman_sherpa_2001}).
\newpage

\appendix 
\section{Principal Component Analysis}
We outline the standard PCA procedure following the seminal work by \cite{murtagh_principal_1987}. Mathematically, PCA is an orthogonal transformation which transforms the data from a basis of variables in which the data is correlated, to a basis in which the data is linearly uncorrelated. Any new data can then be projected onto this new basis.
Consider a set of $N$ spectra covering the same energy range where each spectrum is represented as an $M$-dimensional vector, $X$; \replaced{together, these $N$ vectors span an $M$-dimensional space, $S$}{thus, they generate a M-dimensional vector space, $S$}. The first principal component, $X_0$, is defined to be in the direction of maximum variance in $S$. Subsequently, the $i^{th}$ component is in the direction of the $i^{th}$ highest variance in the perpendicular subspace spanned by the first $i-1$ principal components. There are a total of $M$ principal components. Defining $r_{ij}$ as the initial spectra measurements, where $i$ represents the spectrum's number and $j$ represents the wavelength bin, we can develop the necessary equations:

\begin{equation}
    X_{ij} = r_{ij} - \bar r_i
\end{equation}
\begin{equation}
    \bar r_i = \frac{1}{N}\sum_{i=1}^N r_{ij}
\end{equation}

A covariance matrix, $C$, is constructed in the following fashion,
\begin{equation}
    C_{jk} = \frac{1}{N}\sum_{i=1}^N X_{ij}X_{jk}
\end{equation}
where $1 \leq j,k \leq M$.
The first principal component is defined by the following form:
\begin{equation}
    \mathbf{Ce_1} = \lambda_1\mathbf{ e_1}
\end{equation}
in which $\mathbf{e_1}$ and $\lambda_1$ represent the first (thus numerically largest) eigenvector-eigenvalue pair.
In the context of ICM spectra, these components represent the parameters corresponding to the primary emission variables (temperature, column density, redshift, etc.) or some combination of them. In this work, eigenspectra are calculated using the \texttt{SKLEARN.DECOMPOSITION.PCA} package which uses SVD to decompose the spectra into its principal components.

\section{Weighted Voronoi Tessellation Map of Perseus}
This section includes the weighted voronoi tessellation map of the Perseus cluster and a brief discussion on the signal-to-noise statistics. We see that the WVT bins follow no particular structure as expected. Furthermore, the signal-to-noise of the bins follows a Gaussian distribution with a mean about the target signal-to-noise ratio. 

\begin{figure}
    \centering
    \plottwo{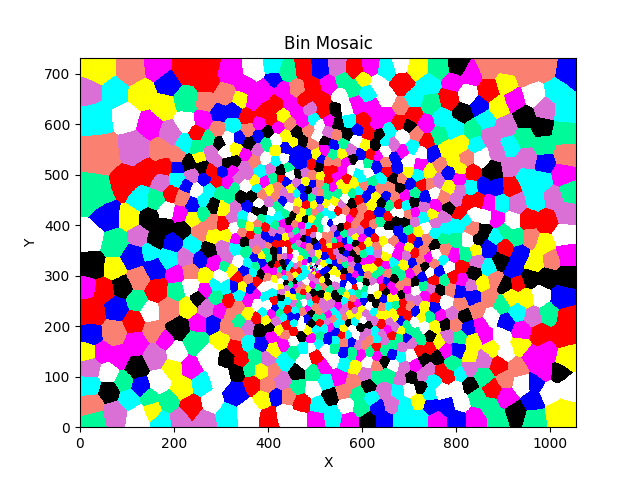}{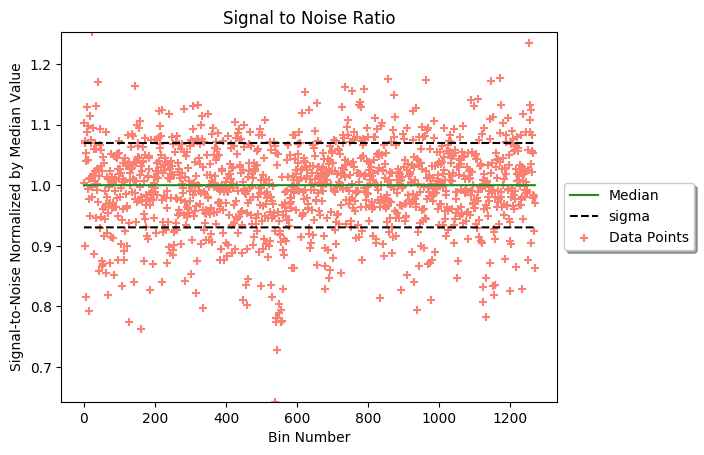}
    \caption{Left: Final bin mosaic of the Perseus cluster. The figure has 1271 separate bins. Each bin is randomly colored from a selection of 10 colors. The X and Y axes are in physical coordinates. Right: Normalized signal-to-noise plot for each bin in the WVT mosaic. We see that the signal-to-noise is tightly constrained around the target signal-to-noise value (150).}
    \label{fig:my_label}
\end{figure}

\section{Component Significance Map}
In this section, we provide the component significance map which shows the relative probabilities of containing x-number thermal components determined by the random forest classifier. We note that the strongest regions of confidence follow the thermal contour shown in figure \ref{fig:Persues_temps}. The existence of this thermal contour has been explored thoroughly in several other papers  (e.g. \citealt{sanders_mapping_2004}; \citealt{sanders_deep_2010}; \citealt{fabian_wide_2011}).

\begin{figure}
    \centering
    \includegraphics[width=0.6\textwidth]{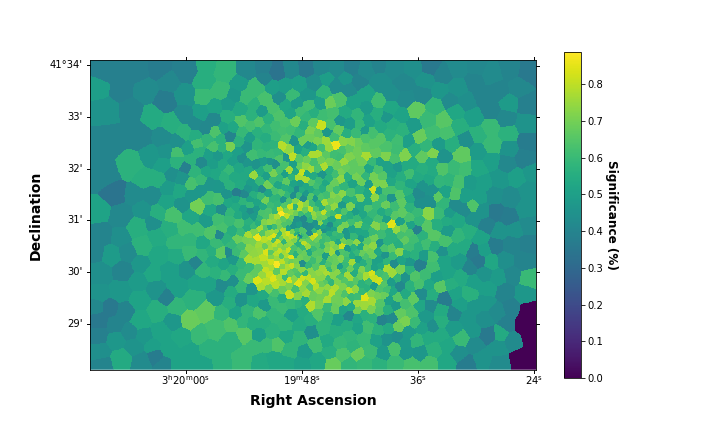}
    \caption{Component significance map of the Perseus Cluster. The value assigned to each pixel represents the probability with which each region was assigned a given number of underlying thermal components. The mean significance is approximately \%70.}
    \label{fig:my_label}
\end{figure}

\bibliography{PCA-Paper}{}
\bibliographystyle{aasjournal}

\end{document}